\documentclass[10pt]{article}

\usepackage{amsmath}
\usepackage{amssymb}

\usepackage{graphicx}

\usepackage{cite}

\usepackage{color} 

\usepackage[labelfont=bf,labelsep=period,justification=raggedright]{caption}

\makeatletter
\renewcommand{\@biblabel}[1]{\quad#1.}
\makeatother

\date{}

\begin{document}

\begin{flushleft}
{\Large
\textbf{ÔNowcastingÕ economic and social data: when and why search engine data fails, an illustration using Google Flu Trends}
}
\\
Paul Ormerod$^{1, \ast}$, 
Rickard Nyman$^{2}$, 
R. Alexander Bentley$^{3}$
\\
\bf{1} Centre for the Study of Decision-Making Uncertainty, University College London, London, United Kingdom
\\
\bf{2} Centre for the Study of Decision-Making Uncertainty, University College London, London, United Kingdom
\\
\bf{3} Department of Anthropology, University of Bristol, Bristol, United Kingdom
\\
$\ast$ E-mail: Corresponding p.ormerod@ucl.ac.uk
\end{flushleft}

\section*{Abstract}

Obtaining an accurate picture of the current state of the economy is particularly important to central banks and finance ministries, and of epidemics to health ministries.  There is increasing interest in the use of search engine data to provide such 'nowcasts' of social and economic indicators.  However, people may search for a phrase because they independently want the information, or they may search simply because many others are searching for it. We consider the effect of the motivation for searching on the accuracy of forecasts made using search engine data of contemporaneous social and economic indicators. We illustrate the implications for forecasting accuracy using four episodes in which Google Flu Trends data gave accurate predictions of actual flu cases, and four in which the search data over-predicted considerably. Using a standard statistical methodology, the Bass diffusion model, we show that the independent search for information motive was much stronger in the cases of accurate prediction than in the inaccurate ones. Social influence, the fact that people may search for a phrase simply because many others are, was much stronger in the inaccurate compared to the accurate cases. Search engine data may therefore be an unreliable predictor of contemporaneous indicators when social influence on the decision to search is strong.  


\section*{Introduction}

The use of search engine data to forecast contemporaneous social and economic indicators has been growing rapidly. \cite{google1, google2} are early examples. \cite{RePEc:nbr:nberwo:19567} note that this type of forecasting, or ÔnowcastingÕ, is of particular interest to central banks, and gives a number of references to such work. In April 2014, the European Central Bank held a two day conference on the use of Big Data in economic policy making, and further examples of search engine use are on the conference website \cite{ecb}.

Obtaining an accurate picture of the current state of the economy is very important to policy makers. Except for financial market data, however, economic data typically appears not only with a lag, but many of the initial estimates are subject to revisions in the future, which can often be substantial.  It has been known since the 1970s that this can cause problems for economic forecasts  \cite{paul1}.

Of course, search engine data has also been used in social contexts, such as health. The large discrepancy between the number of flu cases indicated by Google Flu Trends and the number reported by the Centre for Disease Control and Prevention has been discussed extensively. Two reasons for this are suggested in \cite{vesp}, for example, as being big data hubris and algorithm dynamics.

There is, however, a very general issue with using search engine data to ÔnowcastÕ, whether the series of interest to policy makers is economic or social. This is based on the motivations of the agents when they engage in the search process, a point which has not been addressed in the literature.  People may search for a phrase because they genuinely want the information (ÔindependentÕ searching), or they may search simply because many others are searching for it (ÔsocialÕ searching). 

The overall aim of the work is to examine whether the relative important of independent and social searching has implications for the potential accuracy of ÔnowcastsÕ carried out using search engine data.  

We postulate that if agents are searching primarily independently, and social influence on their behaviour is weak, we might reasonably expect that there will be a straightforward relationship between search activity about a topic and the subsequent actual decision.  So, for example, if the number of searches for Ôvehicle shoppingÕ rises, this should prove a good indicator of a rise in automobile sales  \cite{RePEc:nbr:nberwo:19567}.  However, this is much less likely to be the case when the social motive is strong.  Using examples of predictions made using Google Flu Trends, we show that there is empirical support for this hypothesis.

\section*{Results and Discussion}

For over 40 years, the Bass diffusion model \cite{bass}, cited some 7000 times in the literature, has been widely applied to time series data to estimate the relative strengths of the independent and social motives. An analysis \cite{paul2} applying the Bass model to Google Trends data of the swine flue outbreak of 2009 and those of bird flue in both 2005 and 2009 suggests that the relative weights were different in these different outbreaks.

The Google Flu Trends website (www.google.org/flutrends) contains charts for a number of countries of historical estimates of the incidence flu made from Google Trends data (GT), and official estimates of the number of cases of flu. We identify four clear cases in which the GT data very substantially overestimated the actual number of cases. These are all from developed countries, where we might reasonably expect to rely on the data for flu cases. Namely, in the United States during the winter of 2012/13, Switzerland in 2008/09, Germany 2008/09 and Belgium in 2008/09.
  
As comparators, we take an example from each of these countries of where the GT data appears from the charts to have given an accurate estimate of the number of flu cases. These are for the United States in 2011/12, Switzerland 2007/08, Germany 2005/06 and Belgium 2007/08.

Table 1 shows the adjusted $R^2$, the degree of fit, of the non linear regressions, which in each case is reasonable.
  
The results for the estimated parameters are reported in Table 2. We show the relative size for each country of the estimated independent and social search coefficients in the cases where GT proved correct and those where it was incorrect. In other words, we compare the relative size of the independent coefficients in the cases when GT was correct and when it was incorrect, and similarly for the relative size of the social coefficients.
  
So, for example, for the United States, the estimate value of the independent search coefficient where GT estimates proved correct is considerably greater than it is where GT estimates were substantially incorrect. Similarly, the reverse obtained with the relative estimates of the socially influenced search coefficient.
  
These results are of general importance for the use of search engine data to ÔnowcastÕ social and economic variables of interest to policy makers. Provided that the motivation for search is that which is consistent with standard economic theory, namely that agents are searching independently, the search data can in principle provide accurate, valuable information. However, the more important is social influence in the search behaviour, the less reliable the search engine data may be.
  
An important task is therefore to develop heuristics which can give some indication during the relatively early phase of a rise in search activity of the relative importance of the individual and social motives for the searches. \cite{paul2} show that, in the context of the Bass model, a rapid rise followed by a slow decline indicates that independent motives are important, whereas a more symmetrical outcome shows that the social motive is strong. \cite{paul3} develop a schema for classifying different markets on these lines.
  
We note that the problems identified here, based on the motivations of the agent carrying out the search procedure, will undoubtedly be relevant to other web based platform such as Twitter and Facebook, not least because of the deliberate gaming of such sites noted by \cite{vesp}.  
  
Overall, our results suggest that the use of Big Data per se, illustrated here by search engine data, may not necessarily be useful to policymakers in obtaining accurate pictures of contemporaneous social and economic indicators.  It needs to be used in conjunction with social science theory in order to be able to begin to identify the circumstances in which it may be able to give accurate ÔnowcastsÕ.

\section*{Materials and Methods}

Standard economic theory holds that agents take decisions independently, and have stable tastes and preferences. This remains the case even under bounded rationality, when agents may be subject to the constraint of imperfect information (for example, \cite{akerlof}). However, agents may also be influenced directly by the decisions of others. A classic early reference within economics itself \cite{alchian} suggests that in many circumstances this may be a very sensible course of action to adopt. Copying the recent success of others proved the winning strategy in a recent tournament among social learning algorithms \cite{rendell}.   

Agents may therefore carry out searches using a mixture of these motives.  At one extreme, the search will be purely motivated by the independent desire to obtain information, and the other it will be motivated purely by social influence.  In practice, a mixture of the two will obtain.   

We fit the Bass model to the Google Trends data from the preceding to the following low points either side of the peak level of searches. We model the number of searches for a given week $t$, $S(t)$, by

$$S(t) = m \frac{(p + q)^2}{p}\frac{e^{-(p+q)t}}{(1 + \frac{q}{p}e^{-(p+q)t})^2}$$
where $m$ is the total cumulative number of searches, $p$ is the parameter of independent search and $q$ is the parameter of socially motivated search.

We use the non-linear regression package in the R library nlmrt.  The data is in Table S1.

The use of diffusion curve analysis, such as the Bass model, to identify the relative strengths of independent and social motivations is justified in the context of searches for flu. Potential problems identified in the social learning literature, do not seem relevant in the context of flu searches. \cite{hoppit}, for example, show that a theoretical model based purely on independent motivation may yield patters similar to those generated by diffusion processes with social motivation if, for example, some agents abandon their task at some stage in the process because it is too difficult, or if the impact of neophobia (a fear of the new) diminishes with time.  However, the concept of flu is widely known and understood, and the search process is very simple, so such problems do not seem relevant in the current context.

\section*{Acknowledgments}

The original source for the Google Flu Trend data is the official Google website \\http://www.google.org/flutrends/about/how.html, and the particular data used is set out in Table S1. This work was made possible through Institute for New Economic Thinking Grant No. IN01100025

\bibliography{plos_template}

\section*{Figure Legends}

\section*{Tables}

\begin{table}[!ht]
\center
\caption{
\bf{Adjusted $R^2$ for the non linear regressions of the Bass diffusion model}}
\begin{tabular}{|c|c|c|}
\hline
Country & Correct & Incorrect \\
\hline
United States & 0.937 & 0.852 \\
\hline
Germany & 0.936  & 0.814 \\
\hline
Belgium & 0.806  & 0.930 \\
\hline
Switzerland & 0.926  & 0.958 \\
\hline
\end{tabular}

\label{tab:label}
 \end{table}

\begin{table}[!ht]
\center
\caption{
\bf{Bass model coefficients}}
\begin{tabular}{|c|c|c|}
\hline
Country & Independent search coefficient & Social search coefficient  \\
\hline
United States & correct $\gg$ incorrect & correct $\ll$ incorrect\\
\hline
Germany & correct $>$ incorrect & correct $<$ incorrect\\
\hline
Belgium & correct  $\gg$ incorrect & correct  $\ll$ incorrect\\
\hline
Switzerland &  correct  $\gg$ incorrect & correct  $<$ incorrect\\
\hline
\end{tabular}
\begin{flushleft}The comparative size of Bass model coefficients, representing independent and socially influenced searches, between 'correct' and 'incorrect' forecasts.
\end{flushleft}
\label{tab:label}
 \end{table}

\end{document}